# Investigating the Performances and Vulnerabilities of Two New Protocols Based on R-RAPSE

**Seyed Salman Sajjadi Ghaemmaghami[1], Afrooz Haghbin[2] and Mahtab Mirmohseni[3]**

**[1] Department of Computer Engineering, Science and Research branch,**
**Islamic Azad University, Tehran, Iran**
***Salman.ghaemmaghami@srbiau.ac.ir***

**[2] Department of Computer Engineering, Science and Research branch,**
**Islamic Azad University, Tehran, Iran**
***haghbin@srbiau.ac.ir***

**[3] Department of Electrical Engineering, Sharif University of Technology, Tehran, Iran**
***mirmohseni@sharif.edu***

**Abstract**

Radio Frequency IDentification (RFID) is a pioneer technology which has depicted a new lifestyle for humanity in all around the world. Every day we observe an increase in the scope of RFID applications and no one cannot withdraw its numerous usage around him/herself. An important issue which should be considered is providing privacy and security requirements of an RFID system. Recently in 2014, Cai et al. proposed two improved RFID authentication protocols based on R-RAPS rules by the names of IHRMA and I2SRS. In this paper, we investigate the privacy of the aforementioned protocols based on Ouafi and Phan formal privacy model and show that both IHRMA and I2SRS protocols cannot provide private authentication for RFID users. Moreover, we show that these protocols are vulnerable to impersonation, DoS and traceability attacks. Then, by considering the drawbacks of the studied protocols and implementation of messages with new structures, we present two improved efficient and secure authentication protocols to ameliorate the performance of Cai et al.'s schemes. Our analysis illustrate that the existing weaknesses of the discussed protocols are eliminated in our proposed protocols.

***Keywords:*** *Authentication, RFID protocol, Privacy, Security, Ouafi Phan privacy model, Traceability, Impersonation*

## 1. Introduction

Nowadays, our world is transitioning from an internet of connected individuals to an internet in which everything and everyone is connected, also known as Internet of Things (IoT) [1]. Radio Frequency IDentification (RFID) is a technology which provides a contactless identification through magnetic waves. Health-care, livestock and animal tracking, access control, transportation and supply chain can be mentioned as its applications which play an important roles to prepare the structures for developing the concept of IoT [2-6]. As it is shown in Fig. 1, RFID systems involve three main parts: back-end server, reader and tag. The tag is a microchip which can be attached to different objects with different purposes in an RFID system that falls in one of the three classes: active, passive and semi-active [7]. A passive tag does not have any battery and obtains sufficient energy to reply the reader from the magnetic field achieved through sending the request by the reader. An active tag contains an inner battery, allows it to start a new connection with the reader over than only be a responder. Although the semi-active tag holds an inner battery, it just responds to the received queries from the reader, and performing the internal operations are the only usage of the internal battery [8]. Decreasing the size and cost of RFID tags, have been led to popularity and vast implementation of passive tags in most of novel applications. The back-end server stores all the information of the tags and the readers, and establishes a connection with the tag via tranceiving data with the reader and after investigating the correctness of transferred messages, authenticates the reader and the tag. Although, RFID technology is developing rapidly and providing comfort for users, deficiency of supplying the necessary security, will result in irreparable damages [9]. Therefore, scholars have proposed various type of protocols to provide security and privacy of end-users in RFID systems, which generally classify into four classes based on the deployed cryptographic functions [10]. Full-fledged are the first classes, include ordinary cryptographic functions such as public or private key cryptography systems, one-way hash functions and so forth [2]. Random Number Generators (RNG) and one-way hash functions are permitted to use in the second class. The third class is called lightweight, includes RNG functions and Cyclic Redundancy Code (CRC) checksums[4]. Finally, ultra-lightweight is the last classification, limited to the usage of simple bitwise operators such as AND, OR and XOR [11].

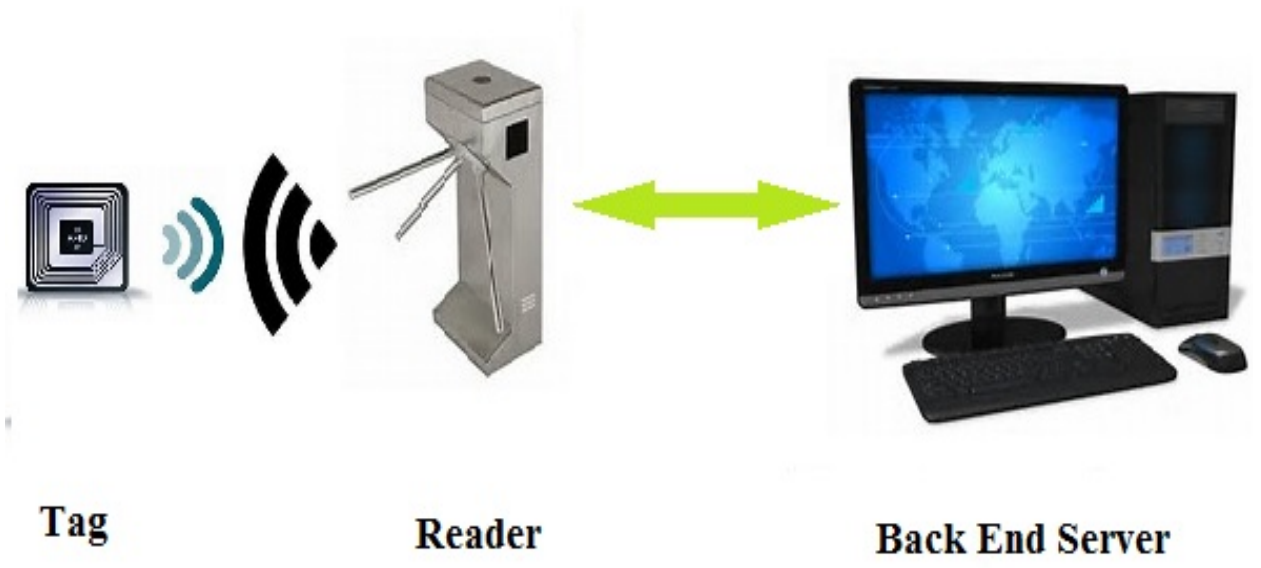

Fig. 1. A System model of RFID systems

By paying attention to the mentioned classification, several protocols have been presented in the last few years [6, 12-16]. Yeh et al. proposed an RFID authentication protocol based on EPC Class 1 Generation 2 standard in 2010 which supplies tag privacy [6]. In 2011, Yoon declared that Yeh et al.'s protocol is still vulnerable to data integrity and forward secrecy problems [13]. So, he suggested an improved protocol available in places with high level of security. In 2011, Cho et al. proposed a hash-based RFID mutual authentication protocol [17]. They believe that their scheme solves the privacy and forgery problems as well as provides all the security requirements of RFID users [17]. In 2014, Cai et al. investigated Cho et al. and Yoon's protocols through their paper [18]. They believe that preparing the authentication procedure should be designed and confirmed theoretically before testing experimentally. So, they defined rules by the name of R-RAPSE which provides security and privacy for RFID protocols [18]. They found that Cho et al.'s protocol does not provide data integrity. Moreover, it is vulnerable to de-synchronization attack, therefore they proposed the IHRMA protocol based on R-RAPSE rules, which is an improved version of Cho et al.'s protocol. They also diagnosed that Yoon's protocol cannot preserve the location of tag owner, which results in weaknesses for providing privacy issues. Thus, they proposed the I2SRS protocol as a modified for Yoon's protocol [7]. In this paper, we analyze the IHRMA and I2SRS protocols and show that there are still some flaws in their protocols. It is exposed that similarity between the generated messages and updating procedures will make the IHRMA protocol vulnerable to tag and reader impersonation. Beside revealing the secret key in the I2SRS protocol, shown that an attacker can perform tag impersonation, traceability and forward traceability attacks after maximum $2^{16}$ computations and backward traceability attack with $2^{17}$ runs in the worst case.

Recently, different types of privacy model have been proposed to study the authentication routine in RFID protocols [19-26]. Among these formal privacy models, Ouafi-Phan privacy model [23] is one of the well-known privacy models which has been proposed in 2008 and due to pertinent queries for different privacy analysis, it has got more attention by researchers [18-22]. We use this privacy model for privacy analysis of the IHRMA and I2SRS protocols. A summary of Ouafi-Phan privacy model and its usage in privacy analysis can be found in our last paper [7].

The rest of this paper is organized as follows: In Section 2 we analyze the IHRMA protocol and its vulnerabilities are discussed. We do the same for the I2SRS protocol in Section 3. Our improved RFID authentication protocols are presented in Section 4. The proposed protocol are compared with some existing ones in the terms of security and privacy in Section 5. Finally, we conclude the paper in Section 6.

## 2. IHRMA protocol

### 2.1 Analyzes of IHRMA protocol

Cai et al. believe that R-RAPSE rules provide sufficient security in their protocol [18]. So, to prevent de-synchronization of secret information in two sides of protocol, they proposed their improved protocol based on R-RAPSE instructions which provide data integrity via hash functions in their method. The structure of IHRMA protocol is depicted in Fig. 2. The connections between the reader and the back-end server are secure, while the communications between the reader and the tag are insecure. In this section, we analyze the IHRMA protocol and prove that their protocol is vulnerable to tag and reader impersonation attacks. The notations used in this protocol are as follow:

*RID*i : Group ID of random number.
*IDk* : The ID of the tag k.
*Sj* : Secret value mutually shared between the server and the tag, used in the jth session.
*Rt* : Random number generated by the tag.
*Rr* : Random number generated by the reader.
*C*1 : Data generated by the tag for authentication.
*C*2 : Blind factor.

| Back-end Server ($ID_k, S_j, S_{j-1}, DATA$) | Reader | Tag ($ID_k, S_j$) |
|---|---|---|
| | 1.1 $Request: R_r$ | |
| | 1.2 $\xrightarrow{R_r}$ | |
| | | 2.1 $C_1 = H(ID_k \oplus R_t \oplus R_r \oplus RID_i)$ |
| | | 2.2 $C_2 = S_j(0:47) \parallel ID_k(48:95)$ |
| 4.1 For each tuple ($ID_k;\ S_j$) | 3.1 $\xleftarrow{C_1, C_2 \oplus R_t, R_r}$ | 2.3 $\xleftarrow{C_1, C_2 \oplus R_t}$ |
| Calculate $C_2^* = S_j(0:47) \parallel ID_k(48:95)$ | | |
| Calculate $R_t^*$ | | |
| Compute $RID_i^*$ | | |
| verify $C_1^* \stackrel{?}{=} C_1$ | | |
| if it does not match | | |
| repeat with tuple ($ID_k;\ S_{j-1}$) | | |
| if it does not match, Reveal the protocol | | |
| 4.2 calculate $R_t^* = C_2 \oplus C_2^*$ | | |
| 4.3 calculate $RID_i^*$ | | |
| 4.4 calculate $C_1^* = H(ID_k \oplus R_t^* \oplus R_r \oplus RID_i)$ | | |
| 4.5 calculate $C_3 = H(C_2 \oplus RID_i)$ | | |
| 4.6 calculate $C_4 = R_t \oplus S_{j+1}$ | 5.1 $\xrightarrow{C_3 \parallel C_4 \parallel C_5}$ | 6.1 calculate $C_3^* = H(C_2 \oplus RID_i)$ |
| 4.7 calculate $C_5 = R_t \oplus H(S_{j+1})$ | | if $C_3^* \neq C_3$ |
| | | reveal the protocol |
| 4.8 $\xrightarrow{DATA \parallel C_3 \parallel C_4 \parallel C_5}$ | | else |
| 4.9 updating | | 6.2 calculate $S_{j+1} = R_t \oplus C_4$ |
| $S_j - 1 \leftarrow S_j \quad S_j \leftarrow S_j + 1$ | | 6.4 calculate $C_5^* = R_t \oplus H(S_{j+1})$ |
| | | if $C_5^* \neq C_5$ |
| | | reveal the protocol |
| | | else |
| | | Server is authenticated |
| | | 6.2 updating $S_j \leftarrow S_j + 1$ |

Fig. 2. IHRMA protocol [18]

## 2.2 Tag impersonation attack

Here now we show that the IHRMA protocol cannot prevent an adversary $\mathcal{A}$ from performing impersonation attacks. Cai et al. have used $mod$ operator in construction of $RID_i$ and they have believed that it will result better performance of their protocol [18]. $RID_i$ is described as follow:

$$RID_i = (R_t - R_t mod S_j + 1)(0:47) \parallel (R_t + S_j - R_t mod S_j)(48:95) \quad (1)$$

**Notion1**: There is a characteristic in the $mod$ operator, which is described here. If we have $a$ and $b$ as two integers and $a < b$, then $a \ mod \ b$ will result in $a$.

Using *Notion 1*, in the IHRMA protocol [18], there will be a situation where $R_t < S_j$, results in

$$RID_i = (R_t - R_t + 1)(0:47) \parallel (R_t + S_j - R_t)(48:95)$$

$$RID_i = (1)(0:47) \parallel (S_j)(48:95) \quad (2)$$

Now, an adversary can perform tag impersonation attack in two phases as follows:

**Learning phase**: The attacker $\mathcal{A}$ sends an execute query $(R, T_0, i)$, obtains $\{R_r, C_1, C_2 \oplus R_t, C_3, C_4, C_5\}$. Then by changing the value of $R_t$, which makes new messages as $C_4$ and $C_5$, prevents the back-end server authentication in the tag side. Therefore, the secret value will not be updated in the tag.

**Attack phase**: In session $(i+1)$, the attacker impersonates him/herself as a trusted tag. To this aim, the adversary performs as follows:

**A.** After receiving $R_r'$ from the reader in the $(i+1)$ session, the attacker $\mathcal{A}$ calculates $C_{1,Fic}$ and $D_{1,Fic}$ with the values stored in the last session, and sends them to the reader.

$$C_{1,Fic} = C_1 \quad (3)$$

$$D_{1,Fic} = C_2 \oplus R_t \oplus R_r \oplus R_r^{i+1} \quad (4)$$

**B.** Reader puts $R_r^{i+1}$ beside $C_{1,Fic}$ and $D_{1,Fic}$, sends them all to the back-end server.

**C.** The back-end server calculates $C_2^*$ and obtains $R_t^{i+1}$ as described below:

$$R_t^{i+1} = D_{1,Fic} \oplus C_2^* \quad (5)$$

**D.** By computing $R_t^{i+1}$, the back-end server calculates $RID_{i+1}$. As we stated before, if $R_t < S_j$, the obtained $RID_{i+1}$ is equal with eq. (1).

**E.** The back-end server, computes $C_1^{*(i+1)}$ and authenticates the tag, if $C_1^{*(i+1)} = C_{1,Fic}$. By considering the situation which $R_t < S_j$ and $R_t^{i+1} < S_j$, $RID_{i+1} = RID_i$ and the back-end server will verify the tag as a legal one, update its secret values and send $C_3^{i+1}$, $C_4^{i+1}$ and $C_5^{i+1}$ to the reader. Ultimately, the back-end server endorses the falsified tag as an allowable one.

**Proof**:

$$\begin{aligned} C_1^{*(i+1)} &= H(ID_k \oplus R_t^{i+1} \oplus R_r^{i+1} \oplus RID_{i+1}) \\ &= H(ID_k \oplus R_t^{i} \oplus R_r^{i} \oplus R_r^{i+1} \oplus R_r^{i+1} \oplus RID_{i+1}) \\ &= H(ID_k \oplus R_t^{i} \oplus R_r^{i} \oplus RID_{i+1}) \\ &= H(ID_k \oplus R_t^{i} \oplus R_r^{i} \oplus RID_{i}) \\ &= C_{1,Fic} \end{aligned} \quad (6)$$

The adversary will be succeed, if the assumptions are correct. For random selection of *Rt* and *Rr*, the success probability of each assumption is ½. The total probability of the above attack is ¼, while it's complexity of is two runs of the protocol.

## 2.3 Reader impersonation attack

As discussed in subsection 2.2 the structure of generating $RID_i$ in the IHRMA protocol which is shown in equation (1), allows the attacker to perform tag impersonation attack. Now, we want to express that this type of producing $RID_i$, will result in reader impersonation attack, with the success probability of 1/4, too. Its implementation can be described as follows:

**Learning phase**: The attacker eavesdrops the $i$th session of the protocol by sending an execute query $(R, T_0, i)$ and obtaining $\{R_r, C_1, C_2 \oplus R_t, C_3, C_4, C_5\}$. Now, to prevent updating secret values, the attacker blocks the 5.1 step of the protocol.

**Attack phase**: The attacker $\mathcal{A}$ acts as a reader and starts a new session by generating $R_r^{i+1}$ randomly and sending it to the tag $T_0$, that leads to reader impersonation attack. The manner of these steps are discussed below:

**1.** The tag did not update its secret values in the last session. Moreover, by considering the conformation of equation (2), the tag will generate a new $R_t^{i+1}$, calculate $C_1^{i+1}$ and send it with $D^{i+1}$ to the counterfeit reader which can be written as,

$$C_1^{i+1} = H(ID_k \oplus R_t^{i+1} \oplus R_r^{i+1} \oplus RID_{i+1}) \quad (7)$$

$$D^{i+1} = C_2^{i+1} \oplus R_t^{i+1} \quad (8)$$

**2.** As the secret values of the tag did not update during the last session, $C_2^{i+1}$ is equal with $C_2^i$. Therefore, the attacker calculates $C_4^{i+1}$ and $C_5^{i+1}$ through the received messages $C_1^{i+1}$, $D^{i+1}$ and the stored values of last session, as follows:

$$\begin{aligned} C_4^{i+1} &= C_2^{i+1} \oplus R_t^{i+1} \oplus C_2^i \oplus R_t^i \oplus R_t^i \oplus S_{j+1} \\ &= R_t^{i+1} \oplus S_{j+1} \end{aligned} \quad (9)$$

$$\begin{aligned} C_5^{i+1} &= R_t^i \oplus H(S_{j+1}) \oplus C_2^{i+1} \oplus R_t^{i+1} \oplus C_2^i \oplus R_t^i \\ &= R_t^{i+1} \oplus H(S_{j+1}) \end{aligned} \quad (10)$$

and sends $\{DATA \parallel C_3^{i+1} \parallel C_4^{i+1} \parallel C_5^{i+1}\}$ to the tag.

**3.** Occurrence the relation of $R_t < S_j$ for two runs of the protocol with the probability of ½ in each time, resulted the tag to authenticate the attacker $\mathcal{A}$ as a legal one and the attacker performs reader impersonation attack with the probability of ¼.

# 3. I2SRS protocol

Cai et al. proposed the I2SRS protocol based on R-RAPSE rules and believe that their protocol will provide all privacy necessity for end-users [18]. Their protocol is depicted in Fig. 3 and the channels between all parts of this system are insecure. The implemented elements in the I2SRS protocol are listed below:

$EPC_s$: Each EPC block with 96 bits, divided to six blocks with 16 bits length. Xoring these six blocks generates $EPC_s$.
$K_i$: Confirmation key stored in the back-end server.
$P_i$: Access key stored in the tag, to authenticate the back-end server.
$\alpha_i$: The back-end server index, stored in the tag to find the corresponding information of the tag in the server

## 3.1 Analyzes of I2SRS protocol

Cai et al. proposed the I2SRS protocol to provide privacy of end-user [18]. They believed, randomness of $\alpha_i$, prepares sufficient properties to prevent an adversary $\mathcal{A}$, performs any attacks on their protocol. But we find that there are still major weaknesses with their protocol.

| Back-end server $(K_0,P_0,\alpha_0,K_n,P_n,\alpha_n,ID_r,EPC_s,DATA)$ | Reader $(ID_r)$ | Tag $(K_i,P_i,\alpha_i,EPC_s)$ |
|---|---|---|
| | 1. Generate $R_r$<br>1.2 $\xrightarrow{R_r}$ | 2.1 Generate $R_t$<br>2.2 $M_1 = P(EPC_s \oplus R_r \oplus R_t) \oplus K_i$<br>2.3 $\beta = R_t \oplus K_i$<br>2.4 $\alpha_i = \alpha_i \oplus R_t$<br>2.5 $\gamma = R_t \oplus P(\alpha_i \oplus K_i)$ |
| | 3.1 $V = H(ID_r \oplus R_r)$<br>3.2 $\xleftarrow{M_1,\beta,\gamma,\alpha_i,R_r,V}$ | 2.5 $\xleftarrow{M_1,\beta,\gamma,\alpha_i}$ |
| 4.1 For each $ID_r$ in server<br>Verify $H(ID_r \oplus R_r) \stackrel{?}{=} V$<br>For each tuple $(EPC_s; K_o, K_n)$<br>4.2 Calculate $I_O = M_1 \oplus K_o$ and $I_n = M_1 \oplus K_n$<br>Verify $I_O \stackrel{?}{=} P(EPC_s \oplus R_r \oplus \beta \oplus K_o)$<br>Verify $I_N \stackrel{?}{=} P(EPC_s \oplus R_r \oplus \beta \oplus K_n)$<br>$x = O/N$<br>4.3 Verify $\beta \oplus K_x \oplus P(\alpha_i \oplus K_x) \stackrel{?}{=} \gamma$<br>4.4 Calculate $R_t^* = \beta \oplus K_x$<br>Verify $\alpha_x \stackrel{?}{=} \alpha_i \oplus R_t^*$<br>4.5 Calculate $M_2 = P(EPC_s \oplus R_t^*) \oplus P_x$<br>4.6 Calculate $Info = DATA \oplus ID_r$<br>4.7 $MAC = H(DATA \oplus R_r)$<br>4.8 $\xrightarrow{M_2,Info,MAC}$ | 5.1 $DATA = Info \oplus ID_r$<br>5.2 Verify $H(DATA \oplus R_r) \stackrel{?}{=} MAC$<br>5.3 $\xrightarrow{M_2}$ | 6.1 Verify $M_2 \oplus P_i \stackrel{?}{=} P(EPC_s \oplus R_t)$<br>6.2 Updating<br>$K_{i+1} \leftharpoonup P(K_i)$<br>$P_{i+1} \leftharpoonup P(P_i)$<br>$\alpha_{i+1} \leftharpoonup P(R_t \oplus R_r)$ |
| 4.9 Updating<br>If $x = N$<br>$K_o \leftharpoonup K_n \leftharpoonup P(K_n)$<br>$P_o \leftharpoonup P_n \leftharpoonup P(P_n)$<br>$\alpha_o \leftharpoonup \alpha_n \leftharpoonup P(R_t \oplus R_r)$<br>Else<br>$\alpha_n \leftharpoonup P(R_t \oplus R_r)$<br>End if | | |

Fig 3. I2SRS protocol [18].

### 3.2 Secret parameter Reveal

Presence of RFID technology as a main part of different systems, compels researchers to present novel protocol which provide security and privacy. An important issue to be considered is preventing an adversary to access secret values, but in this section, we will show that the I2SRS protocol is vulnerable to secret parameter reveal which is described below:

**Learning phase**: An attacker eavesdrops one session of the protocol and blocks the step 5.3. Therefore, it obtains $\{M_1^i, \beta^i, \gamma^i, \alpha_i, {R_r}^i, V, {M_2}^i\}$ while the tag did not update its secret values.

**Attack phase**: The attacker uses stored values from the last session and by paying attention to their similarity, conjectures the secret value $K_i$, with the probability of 1. The attacker computes $Z$ as below:

$$Z = \beta^i \oplus \gamma^i \oplus \alpha_i$$

$$Z = R_t \oplus K_i \oplus \alpha_i \oplus R_t \oplus P(\alpha_i \oplus K_i)$$

$$Z = K_i \oplus \alpha_i \oplus P(\alpha_i \oplus K_i) \qquad (12)$$

$K_i$ is a 16 bit string, so it is one of the element of $\{\mathfrak{K}_1, \mathfrak{K}_2, \dots, \mathfrak{K}_{2^{16}}\}$. The attacker knows the value of $\alpha_i$ through the last session, so he/she is able to obtain $K_i$ after maximum $2^{16}$ runs as below:

$$\begin{gathered} For\ \ 2 < i < 2^{16} \\ Choose\ K_i \in \{\mathfrak{K}_1, \mathfrak{K}_2, \dots, \mathfrak{K}_{2^{16}}\} \\ W = K_i \oplus \alpha_i \oplus P(\alpha_i \oplus K_i) \\ If\ Z = W \\ Return\ K_i\ as\ a\ correct\ secret\ key \\ End \end{gathered} \qquad (13)$$

Therefore, similarity in generating messages through I2SRS protocol, makes it vulnerable to protection of secret parameters. Moreover, lack of precision in updating procedure, allows the attacker $\mathcal{A}$ to calculate $K$, in the $(i + x)$ session by performing $x$ times $PRNG$ of $K_i$.

### 3.3 Tag Impersonation attack

Revealing the secret parameter in the I2SRS protocol, let an adversary $\mathcal{A}$ impersonate a legitimate tag after maximum $2^{16}$ runs, which proved in Section 3.2. The method of applying this attack is as follows,

**Learning phase**: An adversary $\mathcal{A}$ sends an Execute query $(R, T_0, i)$ and obtains $\{M_1^i, \beta^i, \gamma^i, \alpha_i, {R_r}^i\}$. He/she is able to guess $K_i$ with the probability of 1, as discussed in Section 3.2 and obtain $R_t$,

$$R_t^i = \beta^i \oplus K_i \tag{14}$$

**Attack phase**: In the next session, after sending a request and $R_r^{i+1}$ to the tag by the reader, the attacker $\mathcal{A}$ introduces him/herself as a legal tag and the following events take place:

1. The attacker $\mathcal{A}$ generates $\{M_1^{i+1}, \beta^{i+1}, \gamma^{i+1}, \alpha_{i+1}\}$ as below and sends them to the reader,

$$M_1^{i+1} = M_1^i$$

$$\beta^{i+1} = R_t^i \oplus {R_r}^i \oplus {R_r}^{i+1} \oplus K_i$$

$$\alpha_{i+1} = R_t^i \oplus {R_r}^i \oplus {R_r}^{i+1} \oplus \alpha_i'$$

$$\gamma^{i+1} = R_t^i \oplus {R_r}^i \oplus {R_r}^{i+1} \oplus P(\alpha_{i+1} \oplus K_i) \tag{15}$$

As we do not know the value of $EPC_s$, it is not possible to compute $M_1^{i+1}$ unless we use the same message as $M_1^i$ and combine other messages as stated above. In the other words, by calculating $\alpha_i$ as $P(R_t \oplus R_i)$ and replacing $R_t^{i+1}$ with $R_t^i \oplus {R_r}^i \oplus {R_r}^{i+1}$, Messages are prepared.

2. The reader computes $V = H(ID_r \oplus {R_r}^{i+1})$ and sends $\{M_1^{i+1}, \beta^{i+1}, \gamma^{i+1}, \alpha_{i+1}, {R_r}^{i+1}, V^{i+1}\}$ to the back-end server.

3. As the reader is legal, the back-end server authenticates it and to confirm the tag, for each tuple $(EPC_s; K_o, K_n)$, calculates $I_x = M_1 \oplus K_x$. Given that the tag did not update its secret values in the last session, $x$ refers to the old parameters. Then, the back-end server checks if it is equal with $(EPC_s \oplus R_r \oplus \beta \oplus K_x)$ where,

$$M_1 \oplus K_i = P\left(EPC_s \oplus {R_r}^{i+1} \oplus R_t^{i+1}\right) \oplus K_i \oplus K_i$$

$$= P\left(EPC_s \oplus {R_r}^{i+1} \oplus R_t^i \oplus {R_r}^i \oplus R_t^{i+1}\right)$$

$$= P\left(EPC_s \oplus R_t^i \oplus {R_r}^i\right)$$

$$= (EPC_s \oplus {R_r}^{i+1} \oplus \beta^{i+1} \oplus K_i) \tag{16}$$

4. The back-end server checks the equality of $\beta^{i+1} \oplus K_i \oplus P(\alpha_{i+1} \oplus K_i) \stackrel{?}{=} \gamma^{i+1}$ and obtains,

$$\beta^{i+1} \oplus K_i \oplus P(\alpha_{i+1} \oplus K_i)$$

$$= R_t^i \oplus {R_r}^i \oplus {R_r}^{i+1} \oplus K_i \oplus K_i \oplus P(\alpha_{i+1} \oplus K_i)$$

$$= R_t^i \oplus {R_r}^i \oplus {R_r}^{i+1} \oplus P(\alpha_{i+1} \oplus K_i) = \gamma^{i+1} \tag{17}$$

5. Finally, the back-end server authenticates the attacker as a legal tag and calculates $R_t^{i+1}$.

## 3.4 Tag traceability attack

I2SRS protocol does not provide enough privacy issue and it is vulnerable to traceability attack which can be applied by an adversary $\mathcal{A}$, as described below:

**Learning phase**: The attacker sends an execute query $(R, T_0, i)$ and stores the obtained values $\{M_1^i, \beta^i, \gamma^i, \alpha_i, {R_r}^i, V, {M_2}^i\}$ in this session and blocks the 5.3 step of the protocol to prevent the tag's updating procedure.

**Challenge phase**: An attacker $\mathcal{A}$ chooses two fresh tags $T_0$ and $T_1$ and sends a Test query $(T_0, T_1, i+1)$. After choosing $b \in \{0,1\}$ randomly, the attacker takes the tag $T_b$. Then, he/she starts a new session by generating $R_r^{i+1}$ randomly and sending an Execute query $(R, T_b, i+1)$ to the tag $T_b$, which results in obtaining $\{M_1^{i+1}, \beta^{i+1}, \gamma^{i+1}, \alpha_{i+1}\}$.

**Guess phase**: The attacker $\mathcal{A}$ stops the game $G$ and announces $b'$ as his/her guess of $b$ as:

$$b' = \begin{cases} 0 & if \;\; \alpha_i \oplus \beta_i = \alpha_{i+1} \oplus \beta_{i+1} \\ 1 & otherwise \end{cases} \tag{18}$$

As a result, we get:

$$Adv_A^{upriv}(k) = \left|pr(b' = b) - \tfrac{1}{2}\right| = \left|1 - \tfrac{1}{2}\right| = \tfrac{1}{2} \gg \varepsilon \tag{19}$$

**Proof**: By paying attention to the structure of protocol that is depicted in Fig. 3, it is obvious that preventing the tag updates its secret values during the $i$th session, and presence of $R_t$, in both $\alpha$ and $\beta$ messages, allows an adversary to trace the tag after a successful eavesdropping.

## 3.5 Tag forward traceability attack

Providing non-forward traceability means, preventing an adversary from tracking the specific tag. Even, the owner of the tag must not be able to trace his/her tag after giving it over. The I2SRS protocol does not provide forward traceability immunity and the attacker can trace the tag $T_0$, after $N$ runs of the protocol ($\forall N$). The attack is as follows:

**Learning phase**: In the $i$th session of the protocol, an adversary $\mathcal{A}$ sends a Corrupt query $(T_0, K')$ and obtains $\{K_i, P_i, \alpha_i, EPC_s\}$. Weaknesses of updating process in the I2SRS protocol, lets the attacker calculate $K_{i+2}$ by performing two times PRNG on $K_i$.

**Challenge phase**: An attacker $\mathcal{A}$ chooses two fresh tags $T_0$ and $T_1$ and sends a Test query $(T_0, T_1, i)$. After choosing $b \in \{0,1\}$ randomly, the attacker takes the tag $T_b$. Now, the attacker starts a new session by sending an Execute query $(R, T_0, i+2)$ and obtaining $\{M_1^{i+2}, \beta^{i+2}, \gamma^{i+2}, \alpha_{i+2}, {R_r}^{i+2}\}$.

**Guess phase**: The attacker $\mathcal{A}$ stops the game $G$ and after calculating $\{\mathscr{f}, \mathscr{g}, \mathscr{h}\}$, announces $b'$ as his/her guess of $b$ as follows:

$$\mathscr{f} = P(P(K_i^b)) \quad (20)$$

$$\mathscr{g} = \beta^{i+2} \oplus \mathscr{f} \quad (21)$$

$$\mathscr{h} = P(EPC_s \oplus {R_r}^{i+2} \oplus \mathscr{g}) \oplus \mathscr{f} \quad (22)$$

$$b' = \begin{cases} 0 & if \ \mathscr{h} = M_1^{i+2} \\ 1 & otherwise \end{cases} \quad (23)$$

As a result, we have:

$$Adv_A^{upriv}(k) = \left| pr(b' = b) - \tfrac{1}{2} \right| = \tfrac{1}{2} \gg \varepsilon \quad (24)$$

**Proof**: As stated before, the I2SRS protocol suffers from weaknesses in updating technique and transmitting constant messages. The attacker calculates $K_{i+2}$, like mentioned in learning phase, and obtains ${R_t}^{i+2}$ through $\beta^{i+2} \oplus K_{i+2}$. Moreover, constancy of $EPC_s$ in all sessions lets the attacker performs as:

$$\begin{aligned} \mathscr{h} &= P(EPC_s \oplus {R_r}^{i+2} \oplus \mathscr{g}) \oplus \mathscr{f} \\ &= P(EPC_s \oplus {R_r}^{i+2} \oplus \beta^{i+2} \oplus \mathscr{f}) \oplus \mathscr{f} \\ &= P\left(EPC_s \oplus {R_r}^{i+2} \oplus \beta^{i+2} \oplus P\left(P\left(K_i^b\right)\right)\right) \\ &\qquad \oplus P(P(K_i^b)) \\ &= P\left(EPC_s \oplus {R_r}^{i+2} \oplus {R_t}^{i+2}\right) \oplus K_{i+2}^b = M_1^{i+2} \end{aligned} \quad (25)$$

### 3.6 Tag backward traceability attack

Backward traceability immunity assures the owner of the tag that no one will be able to know what happened to him/her before. One of the other vulnerability to the I2SRS protocol in privacy issue is lack of capability to prevent backward traceability attack. Reasons which result in this vulnerability are using the same value for $EPC_s$ in all sessions and predictability of $K_{i+1}$ in each session. An adversary $\mathcal{A}$ performs this attack as follows:

**Learning phase**: An adversary $\mathcal{A}$ sends a Corrupt query $(T_0, K')$ and obtains $\{K_i, P_i, \alpha_i, EPC_s\}$. As discussed in Section 3.2, $K_i$ is a 16 bit string which is one of the elements of $\{\mathfrak{K}_1, \mathfrak{K}_2, \ldots, \mathfrak{K}_{2^{16}}\}$. Therefore, by knowing the updating process of the tag's secret key, the attacker can guess $K_{i-1}$ at maximum runs of $2^{16}$, which is calculated as:

$$\begin{aligned} &For \ \ 2 < i < 2^{16} \\ &\qquad Choose\ K_{i-1} \in \{\mathfrak{K}_1, \mathfrak{K}_2, \ldots, \mathfrak{K}_{2^{16}}\} \\ &\qquad If K_i \stackrel{?}{=} P(K_{i-1}) \\ &\qquad Return\ K_{i-1}\ as\ a\ correct\ secret\ key \\ &End \end{aligned} \quad (26)$$

We have the same relations for finding $P_{i-1}$, which are results of the same inaccuracy in updating procedure. Therefore, the attacker can guess $P_{i-1}$ at maximum runs of $2^{16}$ as calculated:

$$\begin{aligned} &For \ \ 2 < i < 2^{16} \\ &\qquad Choose\ K_{i-1} \in \{\mathcal{P}_1, \mathcal{P}_2, \ldots, \mathcal{P}_{2^{16}}\} \\ &\qquad If P_i \stackrel{?}{=} P(P_{i-1}) \\ &\qquad Return\ P_{i-1}\ as\ a\ correct\ access\ key \\ &End \end{aligned} \quad (27)$$

**Challenge phase**: An attacker $\mathcal{A}$ chooses two fresh tags $T_0$ and $T_1$ and sends a Test query $(T_0, T_1, i)$. After choosing $b \in \{0,1\}$ randomly, the attacker takes the tag $T_b$, sends an Execute query $(R, T_0, i-1)$ and obtains $\{M_1^{i-1}, \beta^{i-1}, \gamma^{i-1}, \alpha_{i-1}, {R_r}^{i-1}\}$.

**Guess phase**: The attacker finishes the game $G$ and proclaims $b'$ as his/her guess of $b$ as follows:

$$b' = \begin{cases} 0 & if \ \ M_2^{i+2} = P(EPC_s \oplus R_t^*) \oplus P_x \\ 1 & otherwise \end{cases} \quad (28)$$

As a result, we have,

$$Adv_A^{upriv}(k) = \left| pr(b' = b) - \tfrac{1}{2} \right| = \tfrac{1}{2} \gg \varepsilon \quad (29)$$

**Proof**: Choosing a correct value for $K_{i-1}$, let the attacker calculates $R_t^*$ in the $(i-1)$th session by computing $\beta^{i-1} \oplus K_{i-1}$. Then, $M_2^{i+2}$ will be computable, if he/she knows the precise value for the access key. Subliminally, the attacker is able to perform backward traceability attack at maximum runs $2 \times 2^{16} = 2^{17}$ computations.

## 4. Improvements of Cai et al.'s protocols

### 4.1 Improvements of the IHRMA protocol

Structure of generating $RID_i$ can be highlighted as the greatest weak point in IHRMA protocol. So, we improve the IHRMA protocol as depicted in Fig 4. By changing the messages $RID_i$ to $H(S_j \oplus R_t)$, the attacker is prevented to access this message. One of the other vulnerability in the IHRMA protocol is the manner of producing and transmitting $C_3$, $C_4$ and $C_5$ messages, which let the attacker leave the protocol unfinished via varying $R_t$. By omitting $C_4$ and $C_5$ messages through our improved protocol and generating $C_3$ as $H(C_2 \parallel RID_i)$, an adversary is not able to perform DoS attack.

| Back-end Server ($ID_k, S_j, S_{j-1}, DATA$) | Reader | Tag ($ID_k, S_j$) |
|---|---|---|
| | 1.1 $Request: R_r$ | |
| | 1.2 $\xrightarrow{R_r}$ | 2.1 $RID_i = H(S_j \oplus R_t)$ |
| | | 2.2 $C_1 = H(ID_k \oplus R_r \oplus RID_i)$ |
| | | 2.3 $C_2 = H(S_j(0:47) \parallel ID_k(48:95))$ |
| 4.1 For each tuple ($ID_k;\ S_j$) | 3.1 $\xleftarrow{C_1, C_2 \oplus R_t, R_r}$ | 2.4 $\xleftarrow{C_1, C_2 \oplus R_t}$ |
| Calculate $C_2^* = S_j(0:47) \parallel ID_k(48:95)$ | | |
| Calculate $R_t^*$ | | |
| Compute $RID_i^*$ | | |
| verify $C_1^* \stackrel{?}{=} C_1$ | | |
| if it does not match | | |
| repeat with tuple ($ID_k;\ S_{j-1}$) | | |
| if it does not match, Reveal the protocol | | |
| 4.2 calculate $R_t^* = C_2 \oplus C_2^*$ | | |
| 4.3 calculate $RID_i^*$ | | |
| 4.4 calculate $C_1^* = H(ID_k \oplus R_t^* \oplus R_r \oplus RID_i)$ | | |
| 4.5 calculate $C_3 = H(RID_i \parallel C_2)$ | | |
| 4.6 $\xrightarrow{DATA \parallel C_3}$ | 5.1 $\xrightarrow{C_3}$ | 6.1 calculate $C_3^* = H(RID_i \parallel C_2)$ |
| 4.7 updating | | if $C_3^* \neq C_3$ |
| $S_j - 1 \leftarrow S_j \quad S_j \leftarrow S_j + 1$ | | reveal the protocol |
| | | else |
| | | Server is authenticated |
| | | 6.2 updating $S_j \leftarrow S_j + 1$ |

Fig. 4. Improved IHRMA protocol

#### 4.1.1 Tag/Reader impersonation resistance

Presence of the $mod$ operator in the IHRMA protocol causes the occurrence of *Notion 1*. In our improved protocol, we generate $RID_i$ via a hash function which prevents the attacker from omitting $R_t$ in $RID_i$. Moreover, the attacker will not be able to substitute $R_t^{i+1}$ with ($R_t^i \oplus R_r^i \oplus R_r^{i+1}$), cause that our new messages are preserved by hash functions.

#### 4.1.2 DoS attack resistance

Presence of $R_t$ in $C_4$ and $C_5$ messages through the IHRMA protocol, let the attacker XOR a random number with them, which prevent the tag authenticate the back-end server and the reader as a legal parts of RFID system. In our improved protocol we generate $C_3$ as $H(RID_i \parallel C_2)$ and by omitting $C_4$ and $C_5$ messages, $C_3$ is the only oracle sends to the tag for authentication. Therefore, an adversary will not be able to alter messages as he/she desires, which provides a secure and assured authentication procedure.

### 4.2 Improved version of I2SRS protocol

Cai et al. [18] believe that Yoon's protocol [13] does not provide the privacy of end-user, so they improved it and proposed the I2SRS protocol. As we described in Section 3, the I2SRS protocol is still vulnerable to different attacks, so this Section, propose a strengthened versions of Cai et al.'s protocol to overcome its weaknesses, which is shown in Fig. 5. Also, the security and privacy analysis of our proposed protocol is provided.

It can be considered that using the $R_t \oplus R_r$ in producing $M_1$, is one of the main weaknesses in the I2SRS protocol, which helps the attacker to calculate $\gamma$ and $\beta$. Moreover, updating procedure is the other weaknesses which result in traceability attacks. Therefore, we improve the I2SRS protocol by substituting $M_1$, $\alpha_i$ and $\gamma_i$ messages as follows,

$$M_1 = P(EPC_s \oplus R_r) \oplus P(R_t) \oplus K_i \tag{30}$$

$$\alpha_i = \alpha_i \oplus R_{new} \tag{31}$$

$$\gamma = R_t \oplus P(\alpha_i \oplus K_i) \oplus P_i \tag{32}$$

$R_{new}$ is a new random number generated by the tag. Moreover, the updating procedure will alter with,

$$K_{i+1} \longleftarrow P(K_i \oplus R_{new}) \tag{33}$$

| **Back-end server** | **Reader** | **Tag** |
| --- | --- | --- |
| $(K_0, P_0, \alpha_0, K_n, P_n, \alpha_n, ID_r, EPC_s, DATA)$ | $(ID_r)$ | $(K_i, P_i, \alpha_i, EPC_s)$ |
| | 1. Generate $R_r$ | 2.1 Generate $R_t$ and $R_{new}$ |
| | 1.2 $\xrightarrow{R_r}$ | 2.2 $M_1 = P(EPC_s \oplus R_r) \oplus P(R_t) \oplus K_i$ |
| | | 2.3 $\beta = R_t \oplus K_i$ |
| | | 2.4 $\alpha_i = \alpha_i \oplus R_{new}$ |
| | | 2.5 $\gamma = R_t \oplus P(\alpha_i \oplus K_i) \oplus P_i$ |
| | 3.1 $V = H(ID_r \oplus R_r)$ | 2.5 $\xleftarrow{M_1, \beta, \gamma, \alpha_i}$ |
| 4.1 For each $ID_r$ in server<br>Verify $H(ID_r \oplus R_r) \stackrel{?}{=} V$<br>For each tuple $(EPC_s; K_o, K_n)$ | 3.2 $\xleftarrow{M_1, \beta, \gamma, \alpha_i, R_r, V}$ | |
| 4.2 Calculate<br>$I_O = M_1 \oplus K_o$ and $I_N = M_1 \oplus K_n$<br>Verify $I_O \stackrel{?}{=} P(EPC_s \oplus R_r) \oplus P(\beta \oplus K_o) \oplus K_o$<br>Verify $I_N \stackrel{?}{=} P(EPC_s \oplus R_r) \oplus P(\beta \oplus K_n) \oplus K_n$<br>$x = O/N$ | | |
| 4.3 Verify $\beta \oplus K_x \oplus P(\alpha_i \oplus K_x) \oplus P_x \stackrel{?}{=} \gamma$ | | |
| 4.4 Calculate $R_t^* = \beta \oplus K_x$<br>compute $R_{new} = \alpha_i \oplus \alpha_x$ | | |
| 4.5 Calculate $M_2 = P(EPC_s \oplus R_t^*) \oplus P_x$ | | |
| 4.6 Calculate $Info = DATA \oplus ID_r$ | 5.1 $DATA = Info \oplus ID_r$ | |
| 4.7 $MAC = H(DATA \oplus R_r)$ | | |
| 4.8 $\xrightarrow{M_2, Info, MAC}$ | 5.2 Verify $H(DATA \oplus R_r) \stackrel{?}{=} MAC$ | 6.1 Verify $M_2 \oplus P_i \stackrel{?}{=} P(EPC_s \oplus R_t)$ |
| 4.9 Updating<br>If $x = N$<br>$K_o \leftharpoonup K_n \leftharpoonup P(K_x \oplus R_{new})$<br>$P_o \leftharpoonup P_n \leftharpoonup P(P_x)$<br>$\alpha_o \leftharpoonup \alpha_n \leftharpoonup P(R_t \oplus R_r \oplus P_x)$<br>Else<br>$\alpha_n \leftharpoonup P(R_t \oplus R_r \oplus P_x)$<br>End if | 5.3 $\xrightarrow{M_2}$ | 6.2 Updating<br>$K_{i+1} \leftharpoonup P(K_i \oplus R_{new})$<br>$P_{i+1} \leftharpoonup P(P_i)$<br>$\alpha_{i+1} \leftharpoonup P(R_t \oplus R_r \oplus P_i)$ |

Fig 5. The improved version of I2SRS protocol.

$$\alpha_{i+1} \longleftarrow P(R_t \oplus R_r \oplus P_i) \quad (34)$$

The improved protocol is depicted in Fig 5. Now, we describe how the applied modifications remove the mentioned weaknesses in Section 3.

### 4.2.1 Secret reveal resistance

We showed that weaknesses in creation of $\alpha_i$ and its updating manner, will result in achieving the correct $K_i$ by the attacker. But in our improved I2SRS protocol, changing $\alpha_i$ to $P(R_t \oplus R_r \oplus P_i)$, prevents the attacker from obtaining $R_t$ and $K_i$.

### 4.2.2 Replay attack resistance

An adversary $\mathcal{A}$ tries to verify, denies or omits the transferred massages through impersonating a legal tag or reader, in replay attacks. In our improved protocol, usage of a new random number, $R_{New}$, and change the structure of produced messages, makes it impossible for an adversary to re-use the eavesdropped ones through last sessions.

### 4.2.3 Impersonation attack resistance

An adversary $\mathcal{A}$ is necessitous of knowing the values of $K_i$, $R_t$ and $EPC_s$ to compute $M_1$, $\beta$ and $\gamma$ messages. By defining a new format for $M_1$ and $\alpha_i$, an adversary is not able to access the secret value of the improved I2SRS protocol. Beside, usage of the last stored $M_1$ is not profitable for him/her in the current session.

## 5. Results and comparisons

### 5.1 Privacy guarantee

The improved I2SRS protocol provides privacy through changing the method of updating procedure. Moreover, it prevents backward traceability, traceability and forward traceability attacks via defining the updated $K_i$ as $P(K_i \oplus R_{new})$.

After the presented analysis, a comparison of the privacy and the security of the improved IHRMA and I2SRS protocols with some similar protocols are presented in Table 1 and 2, respectively. As it

Table 1. Comparison of improved IHRMA protocol with similar ones.

| Feature → / Protocols ↓ | $v_1$ | $v_2$ | $v_3$ | $v_4$ |
|---|---|---|---|---|
| Cho et al. [17] | NO | NO | NO | NO |
| IHRMA [18] | NO | NO | NO | NO |
| Our protocol | YES | YES | YES | YES |

$v_1$: Protection from tag impersonation
$v_2$: Protection from reader impersonation
$v_3$: Prevention DoS attack
$v_4$: Prevention of traceability attack

Table 2. Comparison of improved I2SRS protocol with similar ones.

| Feature → / Protocols ↓ | $W_1$ | $W_2$ | $W_3$ | $W_4$ | $W_5$ | $W_6$ | $W_7$ |
|---|---|---|---|---|---|---|---|
| Yeh et al. [6] | YES | NO | YES | YES | NO | NO | YES |
| Yoon [13] | NO | NO | YES | YES | NO | NO | NO |
| I2SRS [18] | NO | NO | NO | NO | NO | NO | NO |
| Our protocol | YES | YES | YES | YES | YES | YES | YES |

$W_1$: Secret parameter reveal resistance
$W_2$: Protection of replay attack
$W_3$: Protection impersonation attack
$W_4$: Prevention DoS attack
$W_5$: Prevention of traceability attack
$W_4$: Prevention of forward traceability attack
$W_5$: Prevention of backward traceability attack

can be seen, the proposed protocols are protected against numerous attacks such as traceability, backward traceability, forward traceability, impersonation and DoS. Our proposed protocols provide secure and untraceable communications for RFID elements.

## 6. Conclusions

RFID applications are developing in different areas, which provide comfortability, rapidity and accuracy, but an important issue that must be considered is the assurance of a secure and private connection during the communication procedure. This paper investigated the performance and vulnerabilities of two recent authentication protocols proposed based on R-RAPSE rules. Based on Ouafi-Phan formal privacy model, it is shown that both the IHRMA and the I2SRS protocols cannot provide private authentication for RFID users. To enhance the proficiency of these two protocols, we proposed some improvements in this research. This paper proved that the proposed protocols provide required privacy and security against various types of attacks and can solve the drawbacks of the discussed works.